\documentclass[12pt]{article}

\usepackage{amssymb}
\usepackage{amsmath}
\usepackage{amscd}
\usepackage{latexsym}
\usepackage{graphicx}

\usepackage{caption}

\usepackage{cite}

\newcommand{\be}{\begin{equation}}
\newcommand{\ee}{\end{equation}}
\newcommand{\Dlt}{\Delta}

\newcommand{\bfe}{{\bf e}}

\newcommand{\bS}{{\bf S}}

\newcommand{\bB}{{\bf B}}
\newcommand{\bt}{\beta}

\newcommand{\al}{\alpha}
\newcommand{\ra}{\rightarrow}

\newcommand{\gm}{\gamma}
\newcommand{\om}{\omega}
\newcommand{\Om}{\Omega}

\newcommand{\rgl}{\rangle}
\newcommand{\lgl}{\langle}

\begin{document}

\begin{center}

{\Large{\bf Regulating spin dynamics in magnetic nanomaterials} \\ [5mm]

V.I. Yukalov$^{1,2}$ and E.P. Yukalova$^{3}$ }  \\ [3mm]

{\it
$^1$Bogolubov Laboratory of Theoretical Physics, \\
Joint Institute for Nuclear Research, Dubna 141980, Russia \\ [2mm]

$^2$Instituto de Fisica de S\~ao Carlos, Universidade de S\~ao Paulo, \\
CP 369, S\~ao Carlos 13560-970, S\~ao Paulo, Brazil \\ [2mm]

$^3$Laboratory of Information Technologies, \\
Joint Institute for Nuclear Research, Dubna 141980, Russia } \\ [3mm]

{\bf E-mails}: {\it yukalov@theor.jinr.ru}, ~~ {\it yukalova@theor.jinr.ru}

\end{center}

\vskip 1cm

\begin{abstract}

Magnetic nanomaterials can be used in the construction of devices for information 
processing and memory storage. For this purpose, they have to enjoy two 
contradictory properties, from one side being able of keeping for long time 
magnetization frozen, hence information stored, and from the other side allowing 
for quick change of magnetization required for fast erasing of memory and rewriting 
new information. Methods of resolving this dilemma are suggested based on triggering 
resonance, dynamic resonance tuning, and on quadratic Zeeman effect. These methods 
make it possible to realize effective regulation of spin dynamics in such materials 
as magnetic nanomolecules and magnetic nanoclusters.    
\end{abstract}

\vskip 1cm
Magnetic nanomaterials find wide application in the creation of numerous devices 
in spintronics. There are several types of such materials. Among the most known, 
these are magnetic nanomolecules \cite{Barbara_1,Canechi_2,Yukalov_3,Yukalov_4,
Friedman_5,Miller_6,Craig_7,Liddle_8,Rana_9} and magnetic nanoclusters 
\cite{Kodama_10,Hadjipanayis_11,Wernsdorfer_12,Yukalov_13,Kudr_14}.
There exists the so-called magnetic graphene represented by graphene flakes 
containing defects \cite{Terrones_15,Bekyarova_16}, including various magnetic 
defects \cite{Yaziev_17,Enoki_18,Yukalov_19,Yukalov_20}. Trapped atoms, interacting 
through dipolar and spinor forces, form clouds possessing effective spins 
\cite{Griesmaier_21,Baranov_22,Baranov_23,Stamper_24,Gadway_25,Yukalov_26,Yukalov_27}. 
Quantum dots, often called artificial nanomolecules \cite{Birman_28}, also can 
have magnetization \cite{Schwartz_29,Mahagan_30,Tufani_31}. There exist as well 
nanomolecules, such as propanediol C$_3$H$_8$O$_2$ and butanol C$_4$H$_9$OH that, 
although do not have magnetization in their ground state, but can be polarized 
and can keep magnetization for very long times, for hours and months, depending 
on temperature \cite{Yukalov_4,Yukalov_32}. To be concrete, in the present paper 
we consider magnetic nanomolecules and magnetic nanoclusters, although the similar 
consideration is applicable to other nanomaterials.  

Magnetic nanomolecules have the degenerate ground state, when the molecular 
spin can be directed either up or down. These directions are separated by strong 
magnetic anisotropy, with the anisotropy barrier $10-100$ K. Below the blocking 
temperature $1-10$ K, the spin is frozen in one of the directions. The total 
spins of molecules can be different, between $1/2$ and $27/2$.

Magnetic nanoclusters have many properties similar to magnetic nanomolecules. A 
magnetic nanocluster behaves as a magnetic object with a large spin summarizing 
the spins of particles composing the cluster. The number of particles forming 
a cluster can be up to $10^5$. The magnetization blocking temperature is $10-100$ K. 
A cluster radius is limited by {\it coherence radius} that is between $1$ nm to 
$100$ nm. To form a single-domain magnet, a cluster radius has to be not larger 
than the coherence radius, otherwise the sample becomes divided into several 
domains.     
 
Magnetic nanomolecules or nanoclusters, to be used for memory devices, need to 
possess two properties contradicting each other. From one side, in order to keep 
memory for long time, the spin has to be well frozen, which can be achieved with 
strong magnetic anisotropy. But from the other side, in order to quickly change 
the magnetization, which is necessary for memory erasing or for rewriting the 
information content, it is required to have no magnetic anisotropy that hinders 
spin motion. Thus magnetic anisotropy leads to the dilemma: anisotropy is necessary 
for being able to keep well memory, but it is an obstacle for spin regulation. The 
goal of the paper is to suggest ways of resolving this dilemma.

\vskip 2mm
Let us consider, first, a single nanomagnet, either a nanomolecule or a 
nanocluster, whose Hamiltonians are practically the same by form, only with 
different values of the corresponding parameters. The typical Hamiltonian of 
a nanomagnet reads as
\be
\label{1}
\hat H = - \mu_S \bB \cdot \bS - D S_z^2 + E ( S_x^2 - S_y^2 ) \; ,
\ee
where $D$ and $E$ are the anisotropy parameters. The total magnetic field 
\be
\label{2}
\bB = ( B_0 + \Dlt B ) \bfe_z + H \bfe_x + B_1 \bfe_y
\ee
contains an external constant magnetic field $B_0$, and additional magnetic 
field $\Delta B$ that can be regulated, a feedback magnetic field $H$ created 
by a magnetic coil of an electric circuit, and a transverse anisotropy field 
$B_1$. 

The existence of an electric circuit, with a magnetic coil, where the sample 
is inserted to, is the principal part of the setup we suggest. The action of 
a feedback field, created by the moving spin itself, is the most efficient way 
for spin regulation \cite{Yukalov_3,Yukalov_4,Yukalov_33,Yukalov_34,Yukalov_35}. 

Let at the initial moment of time the sample be magnetized in the direction of 
spin up. At low temperature, below the blocking temperature, the spin direction 
is frozen, even if the external magnetic field is turned so that the sample is 
in a metastable state. The feedback field equation, obtained 
\cite{Yukalov_33,Yukalov_34,Yukalov_35} from the Kirchhoff equation has the 
form
\be
\label{3}
 \frac{dH}{dt} + 2\gm H + \om^2 \int_0^t H(t') \; dt' = 
- 4\pi \eta_{res} \; \frac{dm_x}{dt} \;  , 
\ee
where $\gm$ is the circuit attenuation, $\omega$, resonator natural frequency, 
$\eta_{res}\approx V/V_{res}$, filling factor, $V$, sample volume, $V_{res}$, 
the volume of the resonance coil, and the electromotive force is produced by 
the moving average spin
\be
\label{4}
 m_x = \frac{\mu_S}{V} \; \lgl \; S_x \; \rgl \; .
\ee
     
Spin operators satisfy the Heisenberg equations of motion. Averaging 
these equations, we are looking for the time dependence of the average spin 
components
\be
\label{5}
 x \equiv \frac{\lgl S_x \rgl}{S} \; , \qquad
 y \equiv \frac{\lgl S_y \rgl}{S} \; , \qquad 
 z \equiv \frac{\lgl S_z \rgl}{S} \;  .
\ee
The presence of the magnetic anisotropy leads to the appearance in the 
equations of spin motion of the terms binary in spin operators, which need 
to be decoupled. The standard mean-field approximation cannot be applied, 
being incorrect for low spins. We use \cite{Yukalov_35} the {\it corrected 
mean-field approximation}
\be
\label{6}
\lgl \; S_\al S_\bt + S_\bt S_\al \; \rgl = \left( 2 \; - \;
\frac{1}{S} \right) \; \lgl \; S_\al \; \rgl \lgl \; S_\bt \; \rgl
\ee
that is exact for $S=1/2$ and asymptotically exact for large spins $S\gg 1$, 
where we keep in mind $\alpha \neq \beta$. 

To write down the equations of spin motion in a compact form, we introduce 
the notations for the Zeeman frequency $\omega_0$, dimensionless regulated field
$b$, coupling attenuation $\gamma_0$, and the dimensionless feedback field $h$:
\be
\label{7}
\om_0 \equiv -\; \frac{\mu_S}{\hbar} \; B_0 \; ,
\qquad
b \equiv -\; \frac{\mu_S\Dlt B}{\hbar \om_0} \; ,
\qquad
 \gm_0 \equiv \pi \; \frac{\mu_S^2 S}{\hbar V_{res}} \;  ,
\qquad
h \equiv -\; \frac{\mu_S H}{\hbar \gm_0} \; .
\ee
Also, we define the anisotropy frequencies
\be
\label{11}
\om_D \equiv ( 2S-1) \; \frac{D}{\hbar} \; , \qquad 
\om_E \equiv ( 2S-1) \; \frac{E}{\hbar} \; , \qquad 
\om_1 \equiv - \; \frac{\mu_S}{\hbar} \; B_1 \; ,
\ee
and the anisotropy parameter
\be
\label{12}
  A \equiv \frac{\om_D + \om_E}{\om_0} \; .
\ee

Then the spin equations acquire the structure
$$
\frac{dx}{dt} = - \om_0 ( 1 + b - Az) y + \om_1 z \; ,
$$
\be
\label{13}
\frac{dy}{dt} = \om_0 ( 1 + b - Az) x - \gm_0 h z \; , \qquad
\frac{dz}{dt} = 2\om_E x y - \om_1 x + \gm_0 hy  \; , 
\ee
\be
\label{14}
\frac{dh}{dt} + 2\gm h + \om^2 \int_0^t h(t') \; dt' = 4\; \frac{dx}{dt} \; .
\ee
 
As is seen from the equations, the effective Zeeman frequency is
\be
\label{15}
 \om_{eff} = \om_0 ( 1 + b - Az ) \; .
\ee
This frequency is not constant, but depends on time, since the term $Az$ 
varies with time. The spin polarization varies in the range $-1 < z < 1$, 
hence the term $Az$ varies in the interval $[-A,A]$. This implies that the 
detuning is varying and large,
\be
\label{16}
 \frac{\om_{eff} -\om}{\om_0} = b - Az \; .
\ee

Suppose the initial setup is with spin up, $z_0=1$, and the external field 
$B_0$ turns so that the spin direction up corresponds to a metastable state. 
Nevertheless, the spin can be kept in this state for very long time being 
protected by the magnetic anisotropy. Spin reversal can start only when the 
effective Zeeman frequency is in resonance with the resonator natural 
frequency $\om$. However, the large detuning (\ref{16}) does not allow for 
the resonance to occur. 

Assume that we need to reverse the spin at time $\tau$. To start the spin 
motion, we can organize resonance at this time $t = \tau$ by switching on 
the regulated field $b = b(t)$ and setting $b(\tau) = Az_0$, so that at this 
initial time detuning (\ref{16}) be zero. This initial resonance triggers the 
spin reversal because of which this can be called {\it triggering resonance} 
\cite{Yukalov_36}. The reversal of the longitudinal spin polarization for the 
triggering resonance, realized at different times $\tau$, is shown in Fig. 1. 
Frequencies are measured in units of $\gm_0$ and time, in units of $1/\gm_0$. 

Although the triggering resonance quickly initiates spin reversal, but at 
the last stage, when there is no resonance, there appear long tails slowing 
down the reversal process. The reversal would be much faster provided the 
resonance could be kept during the whole process of spin reversal. This can 
be achieved by switching on the regulated field so that to support the 
resonance by varying $b(t)$, 
\begin{eqnarray}
\label{17}
b(t) = \left\{ \begin{array}{ll}
0 ,         ~ & ~ t < \tau \\
A z_{reg} , ~ & ~ t \geq \tau 
\end{array}
\right. \; .
\end{eqnarray}
Then, till the time $t=\tau$ the spin is frozen by the anisotropy. Starting 
from the time $\tau$, the field $b(t)$ is varied by tuning $z_{reg}$ in such 
a way, that $z_{reg}$ be close to $z$, thus diminishing the detuning,
\be
\label{18}
 b(t) - A z(t) = A [\; z_{reg}(t) - z(t) \; ] \ra 0 \; .
\ee
The time dependence of $z_{reg}$ can be defined by the spin dynamics in 
a sample without anisotropy. This method is called {\it dynamic resonance 
tuning} \cite{Yukalov_37}. The spin reversal under this method is faster 
than in the method of triggering resonance, and there are no tails of spin 
polarization, as is seen from Fig. 2.  

\begin{figure}
\begin{minipage}[c]{0.44\linewidth}
\includegraphics[width=\linewidth]{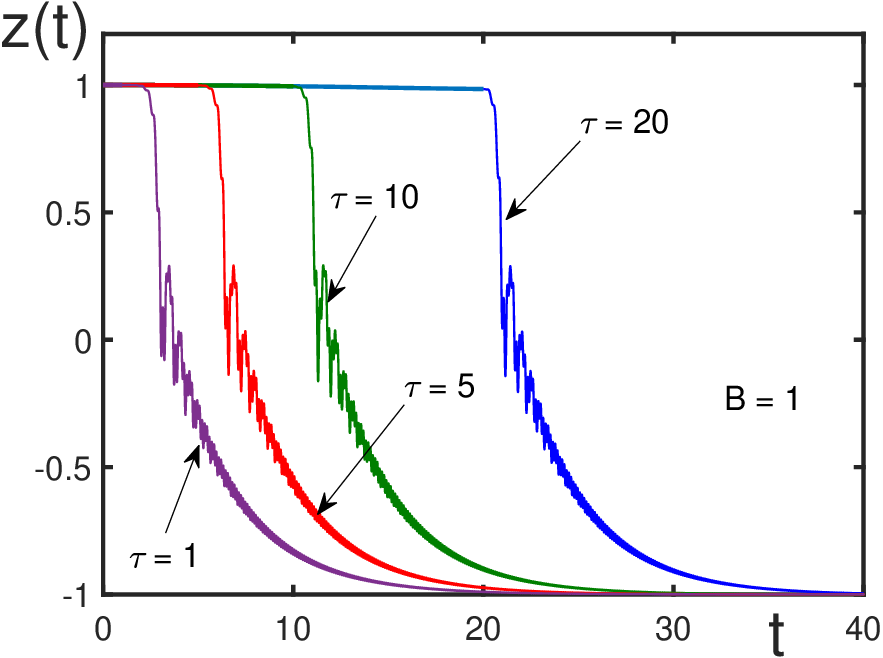}
\caption{\small
Longitudinal spin polarization $z$ as a function of 
dimensionless time $t$ for $\om=\om_0=10$, $\om_E=\om_1=0.01$, $\gm=1$, 
and $b(\tau)\equiv B=Az_0$, with $A=1$ and $z_0=1$. The triggering 
resonance is arranged at different times $\tau$.}
\end{minipage}
\hfill
\begin{minipage}[c]{0.44\linewidth}
\includegraphics[width=\linewidth]{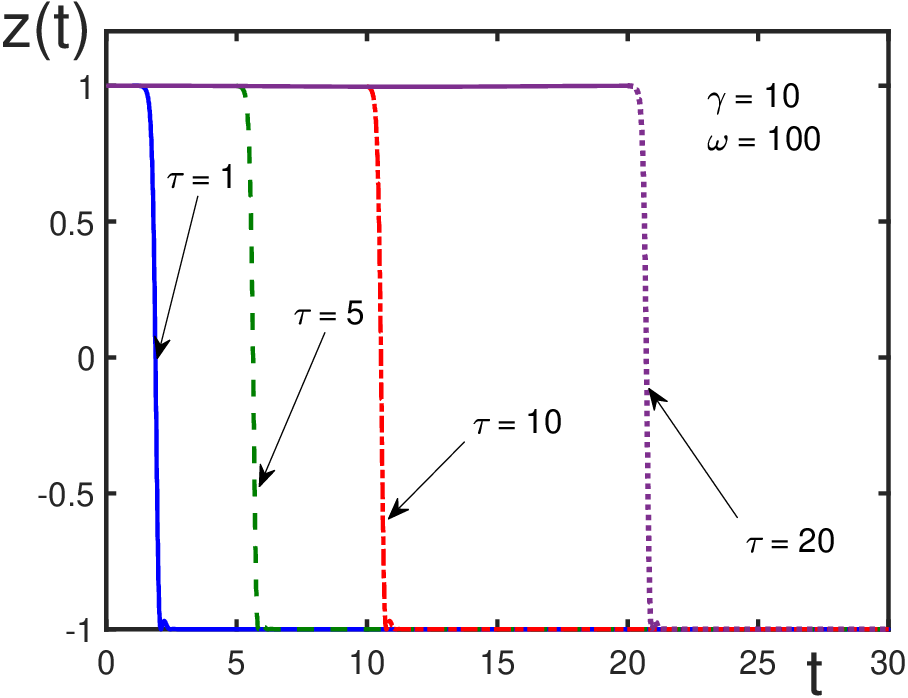}
\caption{\small
Longitudinal spin polarization $z$ as a function of 
dimensionless time $t$ for $\om=\om_0=100$, $\om_E=\om_1=0.01$, $\gm=10$, 
and $A=1$. Dynamic resonance tuning starts at different delay times $\tau$.}
\end{minipage}
\end{figure}

In the case of a sample containing many nanomagnets, it is necessary to 
take into account their interactions through dipolar forces. Then the system 
Hamiltonian
\be
\label{19}
\hat H = - \mu_S \sum_j \bB \cdot \bS_i + \hat H_A + \hat H_D
\ee
contains the Zeeman term, the term of magnetic anisotropy
\be
\label{20}
\hat H_A = - \sum_j D ( S_j^z )^2 \; ,
\ee
and the energy of dipolar interactions
\be
\label{21}
\hat H_D = \frac{1}{2} \sum_{i\neq j} \; \sum_{\al\bt} D_{ij}^{\al\bt}
S_i^\al S_j^\bt \; ,
\ee
where $D_{ij}^{\alpha \beta}$ is a dipolar tensor. The total magnetic field 
is $\bB = ( B_0 + \Dlt B) \bfe_z + H \bfe_x$ .

The methods of triggering resonance and of dynamic resonance tuning regulate 
spin dynamics in a system of many nanomagnets in the same way as in the case 
of a single nanomagnet. The presence of dipolar interactions produces the 
dephasing width $\gm_2$, hence the dephasing time is $1/\gamma_2$. However, 
the spin motion induced by the resonance coupling with a resonant electric 
circuit is coherent and spin reversal happens during the time much shorter than 
the dephasing time. The reversal time can be as short as $10^{-11}$ s. Thus 
dipolar interactions do not hinder the possibility of very fast spin reversal 
when the suggested methods are used. Moreover, under the existence of dipolar 
interactions, there appear dipolar spin waves triggering the initial spin motion 
and facilitating spin reversal \cite{Yukalov_4,Yukalov_33,Yukalov_34,Yukalov_35}.

One more method allowing for the regulation of spin dynamics is based on 
the use of the alternating-current quadratic Zeeman effect 
\cite{Yukalov_26,Yukalov_27,Yukalov_38,Yukalov_39}. Then the Hamiltonian 
for a system of nanomagnets $\hat H = \hat H_Z + \hat H_A + \hat H_D$
contains the same anisotropy term (\ref{20}) and the dipolar term (\ref{21}), 
but the Zeeman term 
\be
\label{24}
\hat H_Z = - \mu_S \sum_j \bB \cdot \bS_j + q_Z \sum_j ( S_j^z)^2
\ee
includes, in addition to the linear part, the quadratic Zeeman term. The 
external magnetic field can be taken in the form $\bB = B_0 \bfe_z + H \bfe_x$.

Writing down the equations of spin motion shows that the effective Zeeman 
frequency becomes $\om_{eff} = \om_0 ( 1 + Az)$, with the effective anisotropy 
parameter
\be
\label{27}
A = ( 2S - 1) \; \frac{q_Z - D}{\hbar\om_0} \; .
\ee

The coefficient of the alternating-current quadratic Zeeman effect 
\be 
\label{28}
q_Z(t) = -\; \frac{\hbar\Om^2(t)}{4\Dlt_{res}(t)}
\ee
can be varied in time by varying the Rabi frequency $\Omega(t)$ and the 
detuning from an internal resonance $\Delta(t)$. By changing the quadratic 
Zeeman-effect coefficient $q_Z(t)$ according to the rule
\begin{eqnarray}
\label{29}
q_Z(t) = \left\{ \begin{array}{ll}
0 , ~ & ~ t < \tau \\
D , ~ & ~ t \geq \tau 
\end{array}
\right. 
\end{eqnarray}
makes it possible to freeze the spin before the time $\tau$ and, when 
necessary, to suppress the effective anisotropy term, thus realizing 
resonance and fast spin reversal. 

In conclusion, we have suggested several ways of regulating spin dynamics 
in magnetic nanomaterials, such as magnetic nanomolecules and magnetic 
nanoclusters. These methods can be employed in spintronics, e.g. for creating 
memory devices.

\end{document}